\documentclass[12pt,preprint]{aastex}

\def\BE{\begin{equation}}
\def\BEL#1{\begin{equation}\label{#1}}
\def\EE{\end{equation}}
\newcommand{\IRAS}{{\it IRAS}}
\newcommand{\COBE}{{\it COBE}}
\newcommand{\WMAP}{{\it WMAP}}

\newcommand{\LPH}{LPH~201.663+1.643}
\newcommand{\HI}{H\,{\scriptsize I}}
\newcommand{\HII}{H\,{\scriptsize II}}

\newcommand{\Halpha}{H$\alpha$}
\newcommand{\etal}{{\it et al.}~}

\newcommand{\degree}{^\circ}

\newcommand{\cm}{{\rm ~cm}}

\newcommand{\GHz}{{\rm ~GHz}}

\begin{document}

\title{Microwave ISM Emission in the Green Bank Galactic Plane
Survey: Evidence for Spinning Dust}

\author{Douglas P. Finkbeiner\footnote{Hubble Fellow, Henry Norris
    Russell Fellow}}
\affil{Princeton University, Department of Astrophysics,
Peyton Hall, Princeton, NJ 08544}

\author{Glen I. Langston, Anthony H. Minter}
\affil{National Radio Astronomy Observatory, P.O. Box 2, Green Bank, WV 24944}


\begin{abstract}
We observe significant dust-correlated emission outside of \HII\
regions in the Green Bank
Galactic Plane Survey ($-4 < b < 4\degree$) at 8.35 and $14.35\GHz$.  The
rising spectral slope rules out synchrotron and free-free emission as
majority constituents at 14 GHz, and the
amplitude is at least 500 times higher than expected thermal dust emission.
When combined with the Rhodes (2.326 GHz), and \WMAP\
(23-94 GHz) data it is possible to fit dust-correlated emission at
$2.3-94$ GHz with only soft synchrotron, free-free, thermal dust, and
an additional dust-correlated component similar to 
Draine \& Lazarian spinning dust.  The 
rising component generally dominates free-free and synchrotron for
$\nu \ga 14\GHz$ and is overwhelmed by thermal dust at $\nu \ga 60\GHz$.
The current data fulfill most of the criteria laid out by
Finkbeiner \etal\ (2002) for detection of spinning dust. 

\end{abstract}



\section{INTRODUCTION}

\label{sec_intro}

The statistical detection of anomalous dust-correlated emission, far
brighter than expected thermal dust emission at
$\sim 10-30\GHz$, has been well established (\COBE, Kogut \etal\ 1996;
Saskatoon, de Oliveira-Costa \etal\ 1997; OVRO, Leitch \etal\ 1997;
19 GHz survey, de Oliveira-Costa \etal\ 1998; Tenerife, de
Oliveira-Costa \etal\ 1999).  There has also been a convincing
detection of this component in the dark cloud LDN1622 (Lynds 1962),
and a detection of a rising spectrum in the diffuse \HII\ region \LPH\ 
(Lockman \etal 1996)
using the Green Bank 140 foot telescope, 
but the interpretation as spinning dust is
tentative (Finkbeiner \etal\ 2002).  In most cases, free-free emission
from ionized gas is ruled
out as the source, either by a significantly rising spectrum or by
morphological comparison with \Halpha\ maps.  This anomalous
dust-correlated emission has been called the ``mystery component'' or
``Foreground X'' (de Oliveira-Costa \etal\ 2002) to avoid premature
interpretation of the signal.  One possible
explanation for this emission is electric dipole emission from rapidly
($\sim 1.5\times10^{10}$s$^{-1}$) rotating small dust grains, known
as ``spinning dust'' (Draine \& Lazarian 1998).  Another possibility
is magnetic dipole emission due to thermal
fluctuations in the magnetization of magnetic grains (Draine \&
Lazarian 1999).  
Henceforth, we follow de Oliveira-Costa and refer to ``Foreground
X,'' keeping these two mechanisms in mind as the most likely. 
Either mechanism might be the
dominant source of dust-correlated microwave emission in the diffuse
ISM at $\sim 15-60\GHz$, overwhelmed by
free-free and synchrotron emission at lower frequencies and
nonmagnetic thermal dust emission at higher frequencies.

Even though Foreground X could be ubiquitous, severe limitations in
previous data sets have prevented a convincing detection over a large
solid angle.  Surveys such as the $19.2\GHz$ survey (Cottingham 1987,
Boughn \etal\ 1992) and the Tenerife \cite{tenerife} survey have such
large beams ($3\degree$ and $5\degree$ FWHM respectively) 
that any diffuse ISM emission in the plane is
confused with \HII\ regions, and even when it can be detected by
spectral slope (as in the Tenerife data at 10 and 15 GHz) there are relatively
few independent pixels in the data showing the rising emission.
Other experiments such as \COBE\ DMR ($7\degree$ FWHM) observed at
higher frequencies where
the spinning dust spectrum is expected to roll off, and cannot be
separated from free-free by spectral slope, or spatial resolution.
This has made it extremely difficult to argue for an unambiguous
detection.

Recently, the \emph{Wilkinson Microwave Anisotropy Probe} (\WMAP\ ; Bennett
\etal\ 2003a) has produced high-sensitivity full-sky maps at 23, 33,
41, 61, \& 94 GHz from its first year of
data\footnote{http://lambda.gsfc.nasa.gov}.  The main purpose of the
mission is to determine cosmological
parameters from CMB anisotropy, and this purpose has already been
largely fulfilled (Hinshaw \etal 2003a, Spergel \etal 2003).
In addition, the \WMAP\ data
provide full-sky maps of Milky Way emission in a frequency domain
seldom explored before.  To interpret this new foreground
information, the \WMAP\ team performed a Maximum Entropy Method (MEM)
analysis as well as more conventional template fitting, and found that
the Foreground X is not necessary to fit the \WMAP\ data,
instead relying on a hard synchrotron component ($T_b \sim\nu^{\beta}$ with
$\beta\sim-2.5$, typical of supernova remnants) that is
strongly correlated with dust (Bennett \etal 2003b).  The spectrum
of this
hard synchrotron component (for $23\GHz<\nu<94\GHz$) is inconsistent
with the Draine \& Lazarian cold
neutral medium (CNM) model chosen for comparison, but does not
disagree strongly with a warm neutral medium (WNM) model scaled
slightly in amplitude.  Indeed, Finkbeiner (2004) demonstrates in a
companion paper that the \WMAP\ data off the plane\footnote{
where \Halpha\ extinction is less than 2 mag according to the Schlegel
\etal (1998) dust map}
can be fit by the following 4
components: 1) soft synchrotron, as traced by 408 MHz (Haslam \etal\ 1982); 2)
free-free, as traced by \Halpha\ emission in three recent surveys 
(VTSS, Dennison \etal\ 1998;
SHASSA, Gaustad \etal\ 2001;
WHAM, Haffner \etal\ 2003) 
as presented by Finkbeiner (2003) plus an enhancement from hot
gas within $30\degree$ of the Galactic center; 3)
Rayleigh-Jeans thermal (vibrational) dust emission as predicted by
Finkbeiner, Davis, \& Schlegel (1999; hereafter FDS), and 4) a Foreground X
template constructed from the FDS dust map times dust temperature
squared.  The spectral shape of each of the four components is
not free to vary with position on the sky, as it is in the \WMAP\ MEM
analysis, yet the fit is superb.
The spectrum derived for Foreground X
can be explained by a superposition of Draine \& Lazarian spinning
dust models, but could also be explained as hard synchrotron highly
correlated with dust.

Lagache (2003) has correlated the \WMAP\ data with \HI\ maps and
found evidence of microwave emission in excess of that expected for
synchrotron, free-free, and thermal dust.  Banday \etal\ (2003) also
address this excess in a new analysis of the \COBE\ data and compare
to the \WMAP\ findings.  Recent measurements of the Helix planetary
nebula at $26-36\GHz$ with the Cosmic Background Imager also
show a factor of 3 excess over the flux expected in the absence of
Foreground X (Casassus \etal\ 2004).  Finally, de Oliveira-Costa
\etal\ (2004) cross-correlate the \WMAP\ MEM synchrotron template
(Bennett \etal\ 2003b) with Tenerife data at 10 and 15 GHz, finding
the correlation spectrum departs from the expected synchrotron shape
by a factor of $\sim10$ at 10 GHz.  Unfortunately, all of these
investigations have some shortcomings: none of the \WMAP-based
analyses have the power to rule out hard synchrotron as the Foreground X
component, because at least some spinning dust models are degenerate
with hard synchrotron over the frequencies observed ($\nu>23\GHz$),
leaving the interpretation of \WMAP\ ambiguous.  A rising spectrum
observed at lower frequencies, seen by de Oliveira-Costa \etal\
(2004), is stronger evidence, but the Tenerife data have a $5\degree$
FWHM resolution.  Fortunately, higher resolution data at appropriate
frequencies do exist. 

The Green Bank Galactic Plane Survey (Langston \etal 2000) at 8.35 and
14.35 GHz provides key evidence in favor of Foreground X, even though it
was designed to detect transient point sources in the
Galactic plane $(|b| < 5\degree)$, not to examine the diffuse
ISM.  The scan pattern is simpler than would be desired for a diffuse
ISM experiment, and 
in the original analysis the data stream was median filtered on a short
timescale to reveal the point sources.  However,
the survey has the requisite resolution, sensitivity, and frequency
coverage to make a highly significant detection of Foreground X.
After careful reprocessing of the raw survey data, we have been able
to produce images of this anomalous emission on a large angular scale
and demonstrate that it is consistent with Draine \& Lazarian spinning
dust, ruling out hard synchrotron as a majority component at 14
GHz and $|b| < 2\degree$. 

In this paper, we compare Foreground X to spinning dust because of
theoretical bias (we know the small grains are there, and they have to
spin and radiate at some level) keeping in mind that alternative
interpretations, such as magnetic dipole emission, are possible.

In \S2 we briefly describe the survey data.  Section 3 describes the
processing steps taken to produce maps of each ``segment'' of sky. 
Results based only on Green Bank data, using the full resolution
($11.2'$) available,
are presented in \S4.  Additional data from Rhodes and \WMAP\ (smoothed
to $1\degree$ FWHM) are analyzed in \S5, and conclusions stated in
\S6.

\section{THE DATA}

The Green Bank Earth Station (GBES) is a 13.7m dish equipped to
communicate with satellites via two receivers centered at 8.35
(X-band) and
14.35 GHz (Ku-band).  These receivers view a common beam center on the sky (to
within 1/8 beam) through a 
dichroic element, allowing simultaneous observation at two
frequencies.  This simultaneity is advantageous for both satellite
communication and transient source surveys, but is also convenient for
our purposes.  The telescope resolution for point sources is
$9.7 \pm 0.1'$ at 8.35 GHz and $6.6\pm0.2'$ at 14.35 GHz.  After
convolving the data samples on to a grid, the effective resolution is 
$11.2'$ at $8.35\GHz$ and $8.0'$ at $14.35\GHz$.  Further details are given in
Langston \etal\ (2000). 

The scan strategy is to raster the telescope through $-5<b<5\degree$
along a line of (nearly) constant Galactic longitude, sampling total
power (both circular polarizations) in a 500 MHz bandwidth 9 times per
second (every $2.4'$)
along the scan.  Successive scans are spaced $5'$ apart, Nyquist
sampling the sky at X-band but not at Ku-band.  Data are
taken in ``segments'' of 180 scans, covering a patch of
$10\degree\times15\degree$ on the sky.  The segments are centered on integer
multiples of $15\degree$ Galactic longitude.

In order to find transient sources, all segments centered on $0
<l<270\degree$ were observed many times and compared.  There are four
complete surveys (GPA, GPB, GPC, and GPD) each of which nominally
consists of 4 epochs a few days apart.  Sometimes a given segment is
observed several times within one epoch to compensate for bad
weather. 
For the current project we use only GPA, which provides up to 5
observations of each segment, though in practice we choose to use the
best three. 

\begin{figure}[tb]
\epsscale{0.8}
\plotone{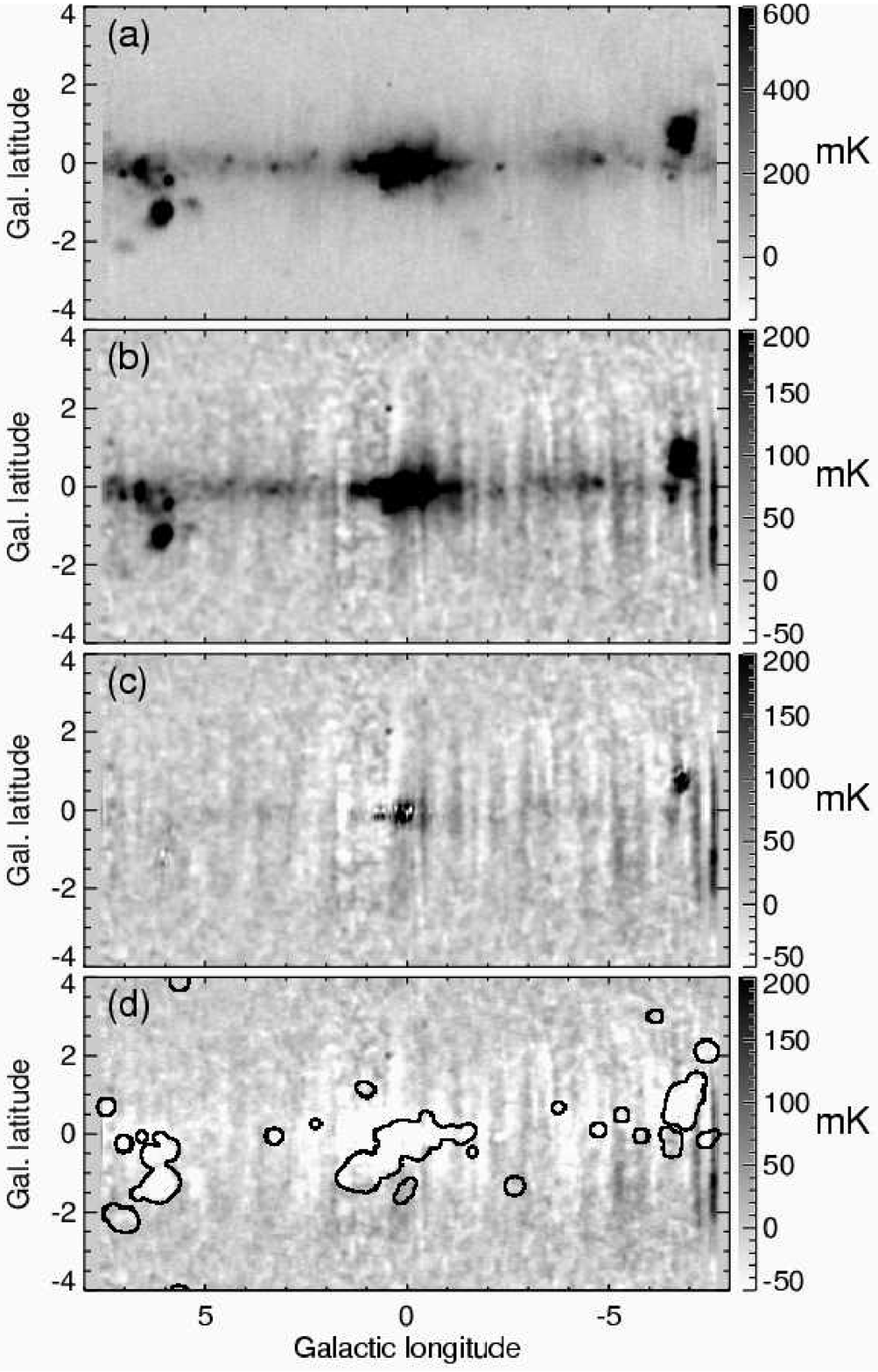}
\figcaption{GPA segment centered on $l=0\degree$: 
$(a)$ X-band (8.35 GHz) map;
$(b)$ Ku-band (14.35 GHz) map, smoothed to X-band beam;
$(c)$ free-free subtracted Ku-band map,
i.e. $T_{Ku}-(14.35/8.35)^{-2.1}T_X$; and
$(d)$ $T_{Ku}-(14.35/8.35)^{-1.2}T_X$ map with outline of \HII\ mask
(see \S \ref{sec_mask})
superposed.  The \HII\ regions in $(d)$ are oversubtracted, but the diffuse
emission is canceled.  This segment exhibits more striping
artifacts than the others.
\label{fig_seg00}
}
\end{figure}
\begin{figure}[tb]
\epsscale{0.8}
\plotone{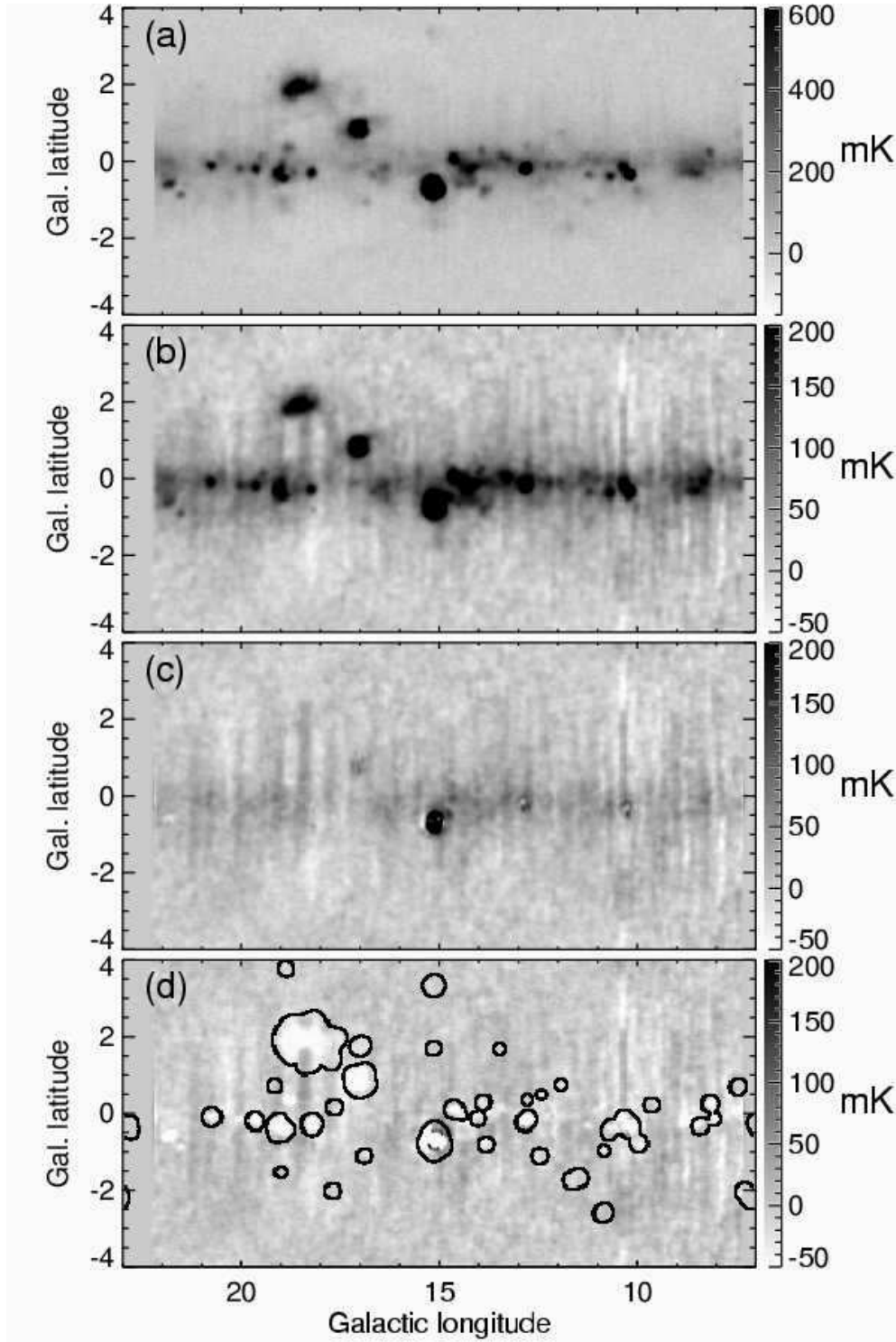}
\figcaption{GPA segment centered on $l=15\degree$, see Figure 1 caption.
The subtraction is good in (c) considering that the region around NGC
6618 $(l,b) = 15.1\degree, -0.7\degree$ is 10 times brighter than the
region around NGC 6611 at $(l,b) = 17.0\degree, 0.9\degree$. 
\label{fig_seg15}}
\end{figure}
\begin{figure}[tb]
\epsscale{0.8}
\plotone{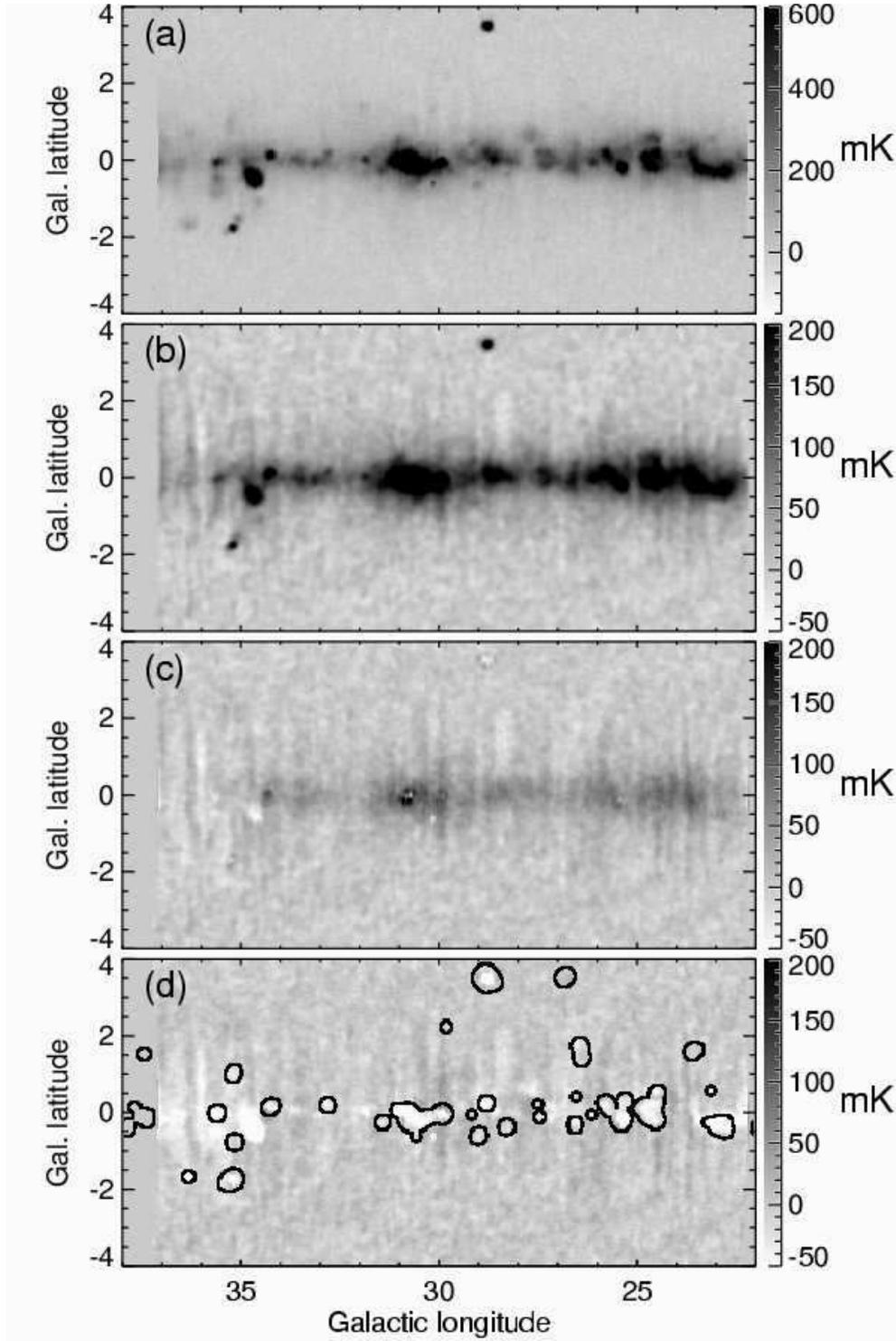}
\figcaption{GPA segment centered on $l=30\degree$; see Figure 1 caption.
Oversubtracted regions in this map are dominated by synchrotron
emission, such as the SNR 3C392 at $(l,b) = 34.75\degree, -0.5\degree$.
The W40 complex is visible at $(l,b) = 28.8\degree, 3.5\degree$, and
is used as a calibration check in \S\ref{sec_calcheck}.
\label{fig_seg30}}
\end{figure}
\begin{figure}[tb]
\epsscale{0.8}
\plotone{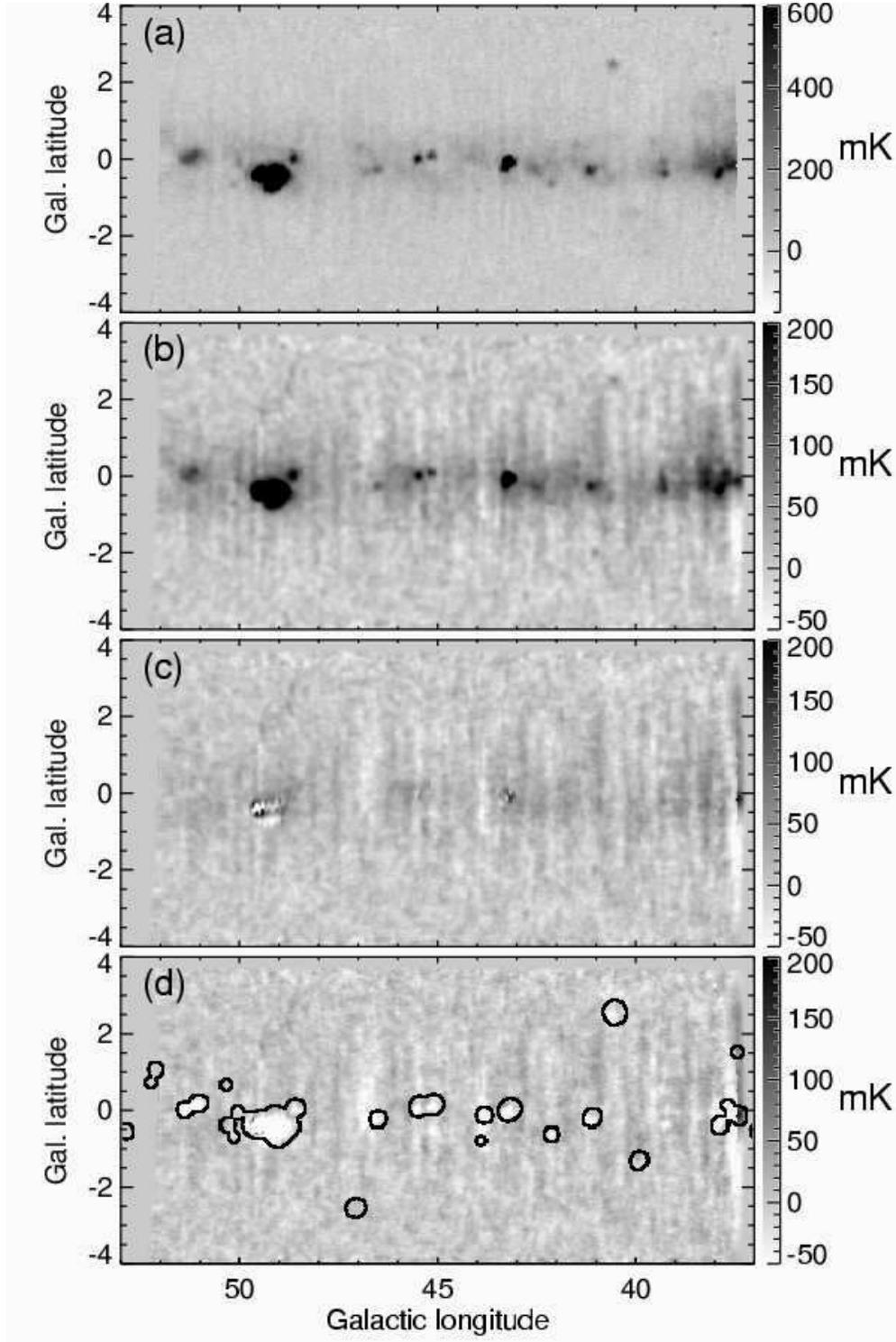}
\figcaption{GPA segment centered on $l=45\degree$; see Figure 1 caption.
\label{fig_seg45}}
\end{figure}

\begin{figure}[tb]
\epsscale{0.9}
\plotone{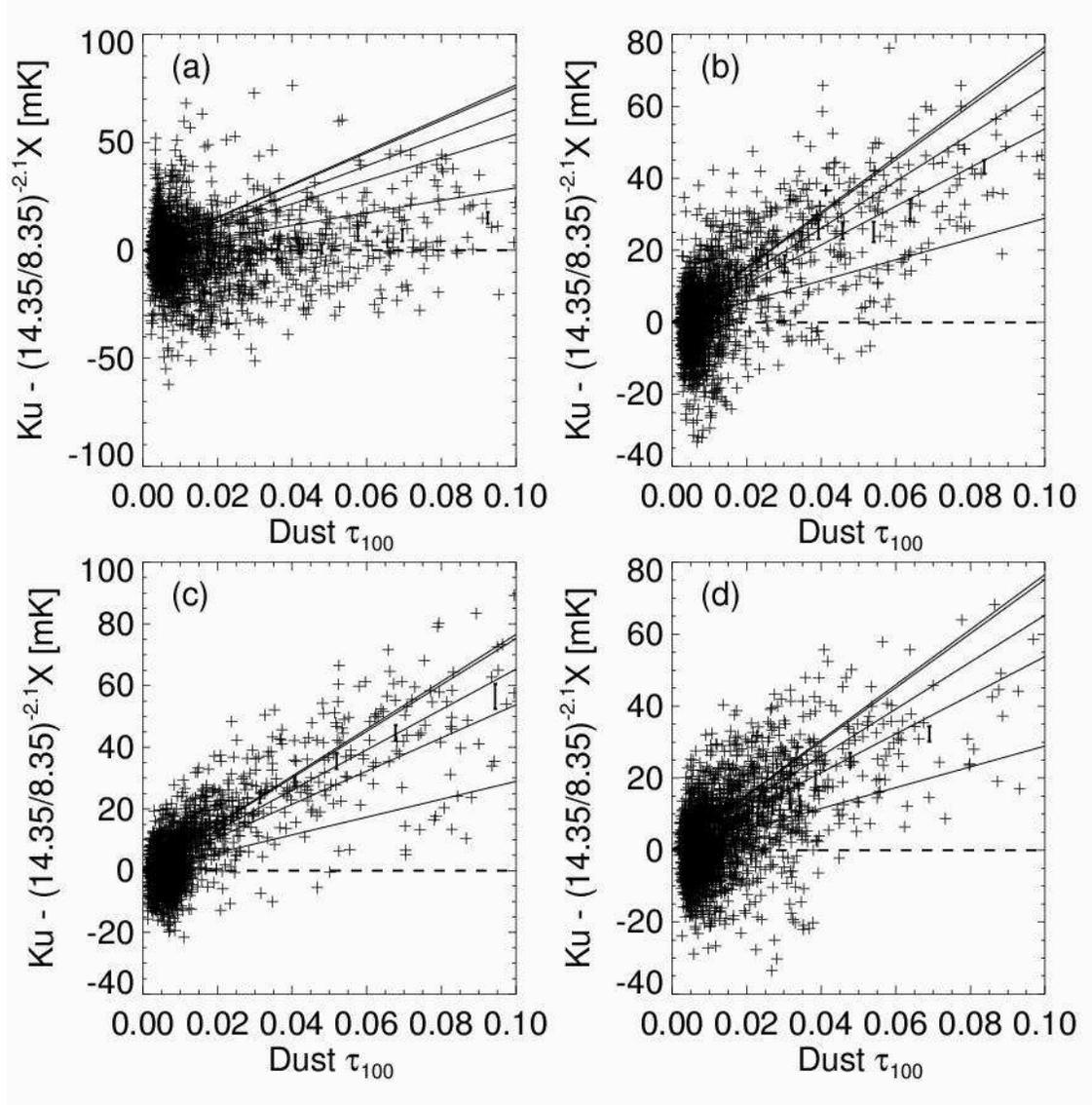}
\figcaption{Free-free subtracted Ku-band antenna temperature
vs. dust optical depth from SFD98 for segments centered at 
($a$) $0\degree$, ($b$) $15\degree$, ($c$) $30\degree$, and ($d$) $45\degree$
Galactic longitude.  Each
symbol represents a $12'\times12'$ patch of sky (independent beams).
Pixels dominated by free-free would fall on the dashed horizontal
line and synchrotron would fall below the line in the absence of noise.
The solid lines represent 5 Draine \& Lazarian (1998) spinning
dust models for (\emph{from top to bottom}) WIM, WNM, CNM, molecular
clouds, and dark clouds, respectively.  Error bars show $1\sigma$
uncertainty of the mean values in bins containing 1/50 of the data
points.  These error bars do not include an overall
calibration uncertainty of 10\%.  Segments 15, 30, and 45 are
consistent with spinning dust models and inconsistent with free-free or
synchrotron emission. 
\label{fig_scatter}
}
\end{figure}
\begin{figure}[tb]
\epsscale{0.9}
\plotone{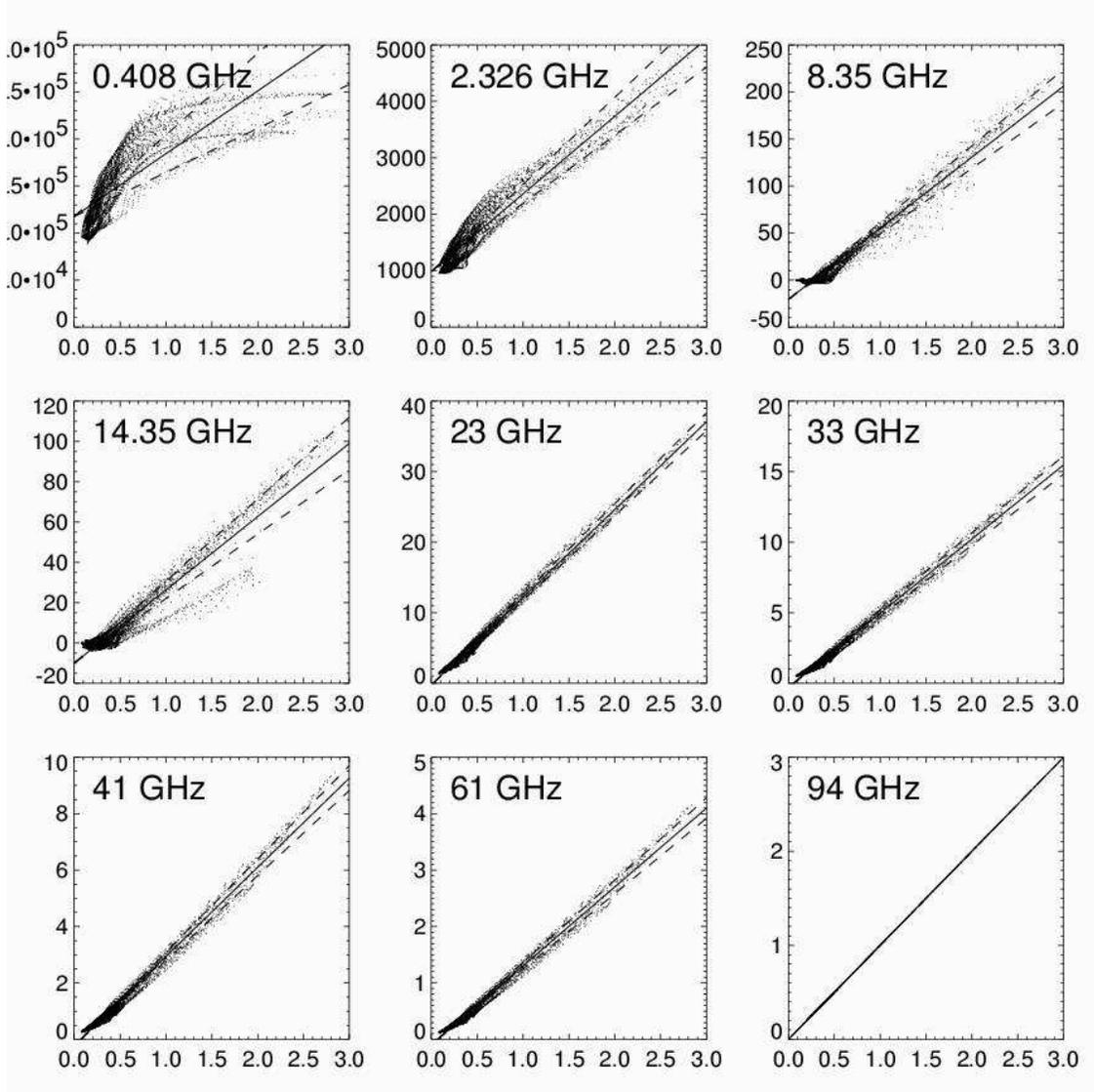}
\figcaption{
Scatter plots for Segment 30.  At each of 9 frequencies, antenna
temperatures [mK] are plotted vs. the \WMAP\ 94 GHz channel, used as a
tracer of dust.  CMB
anisotropy is removed from the 5 \WMAP\ bands before plotting.  When 
this plot is made using e.g. SFD98 $100\micron$ optical depth as a dust tracer, the relative
slopes are the same,
but the scatter is greater.  \WMAP\ 94 GHz is dominated by
Rayleigh-Jeans dust emission, and is therefore a good proxy for total
dust column density.  Scatter plots like these were made for all 4
segments, and the correlation slopes are plotted vs. frequency in
Figure \ref{fig_spectrum4}.  We discard the Haslam 408 MHz result
because the scatter is so non-linear. 
\label{fig_scatter9}
}
\end{figure}
\begin{figure}[tb]
\epsscale{1.0}
\plotone{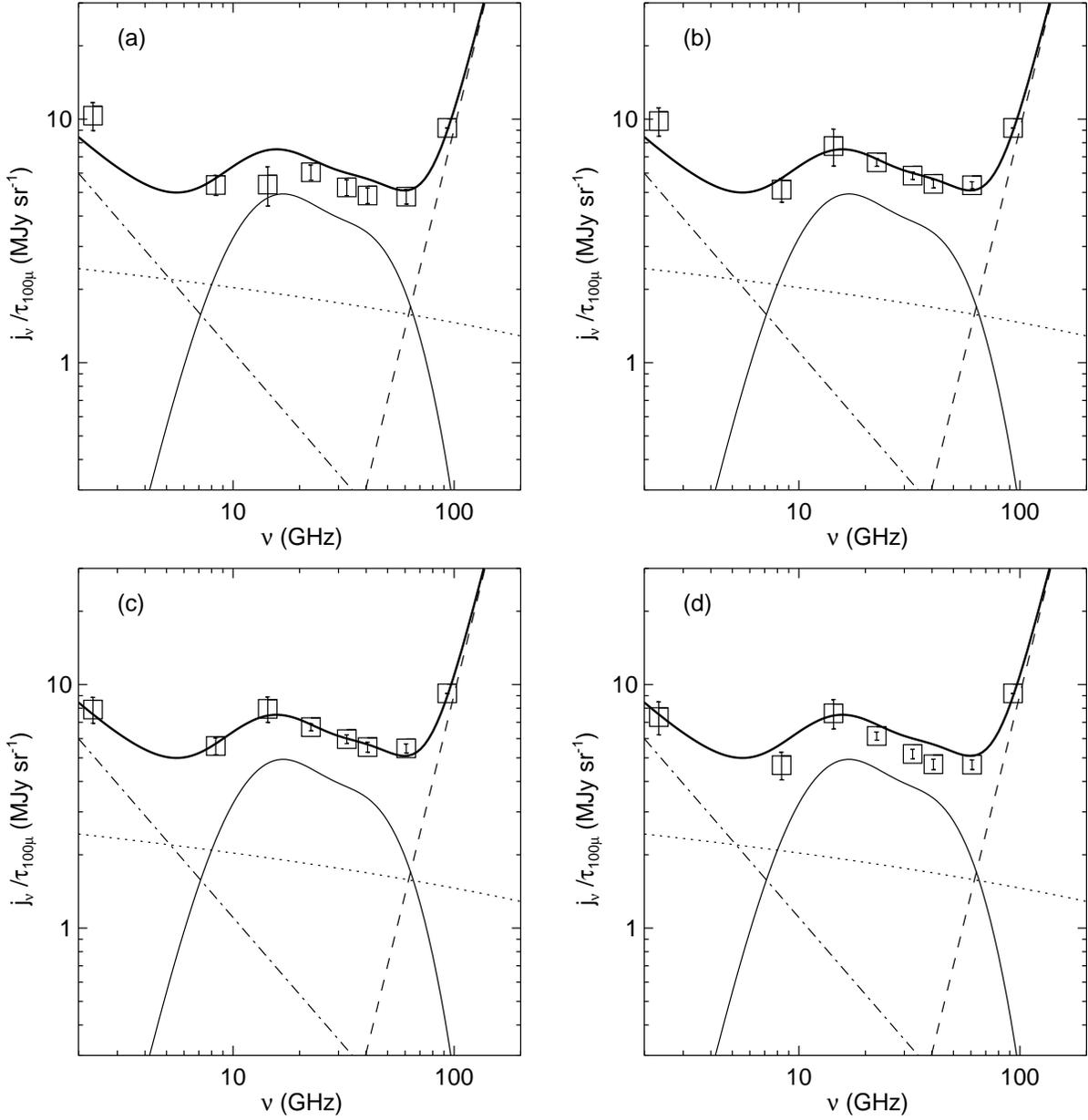}
\figcaption{
Correlation slope spectra for segments $(a)$ 0, $(b)$ 15, $(c)$ 30,
and $(d)$ 45.  The rise from 8 to 14 GHz cannot be explained with any
combination of 
free-free, synchrotron, and thermal dust.  The 94 GHz map is used
as a tracer of dust column density, and is placed on this plot using a
conversion factor calculated in \S\ref{sec_corrslope}.  Also shown are
free-free (\emph{dotted line}), synchrotron with index $\alpha=-1.05$
(\emph{dash-dot line}), thermal dust with Finkbeiner \etal\ (1999) calibration
(\emph{dashed line}), and a mock spinning dust model described in 
\S\ref{sec_corrslope} (\emph{lower solid curve}), 
as well as the sum of these four components (\emph{thick solid curve}).  
The same curves are overplotted in each panel to guide the eye, and
are not a formal fit to the data. 
\label{fig_spectrum4}
}
\end{figure}
\begin{figure}[tb]
\epsscale{0.5}
\plotone{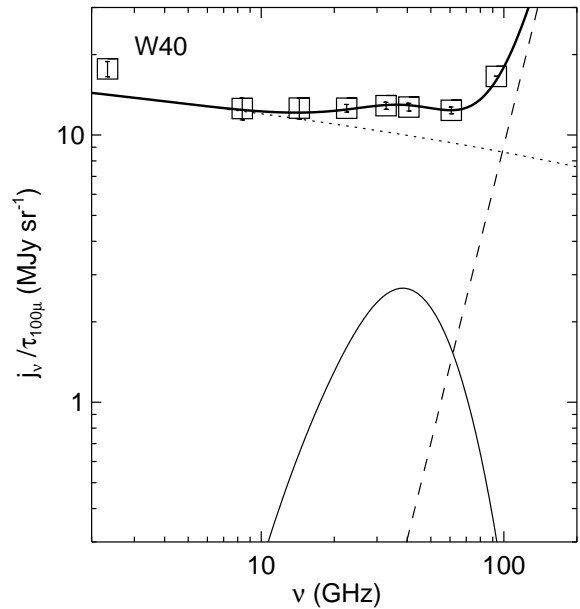}
\figcaption{
Correlation slope spectrum of W40 (also known as LBN90).  W40 is an
\HII\ region around 6 or more OB stars eating into a large molecular
cloud.  The dominance of free-free and distance from the Galactic
Plane ($3.5\degree$) allow this source to be used as a calibration
check.  The curve through the data points is a sum of free free
(\emph{dotted line}), thermal dust (\emph{dashed line}), and 0.15
times the
Draine \& Lazarian model for molecular cloud spinning dust 
(\emph{solid line}). 
This is close to the factor of 0.2 proposed by Draine \& Lazarian
(1998) to account for the expected depletion of small grains in
molecular clouds, but
this plot should not be over interpreted; the point is that the Green
Bank 8.35 and 14.35 GHz data and the \WMAP\ data have a consistent
calibration.
\label{fig_W40}
}
\end{figure}

\begin{figure}[tb]
\epsscale{0.75}
\plotone{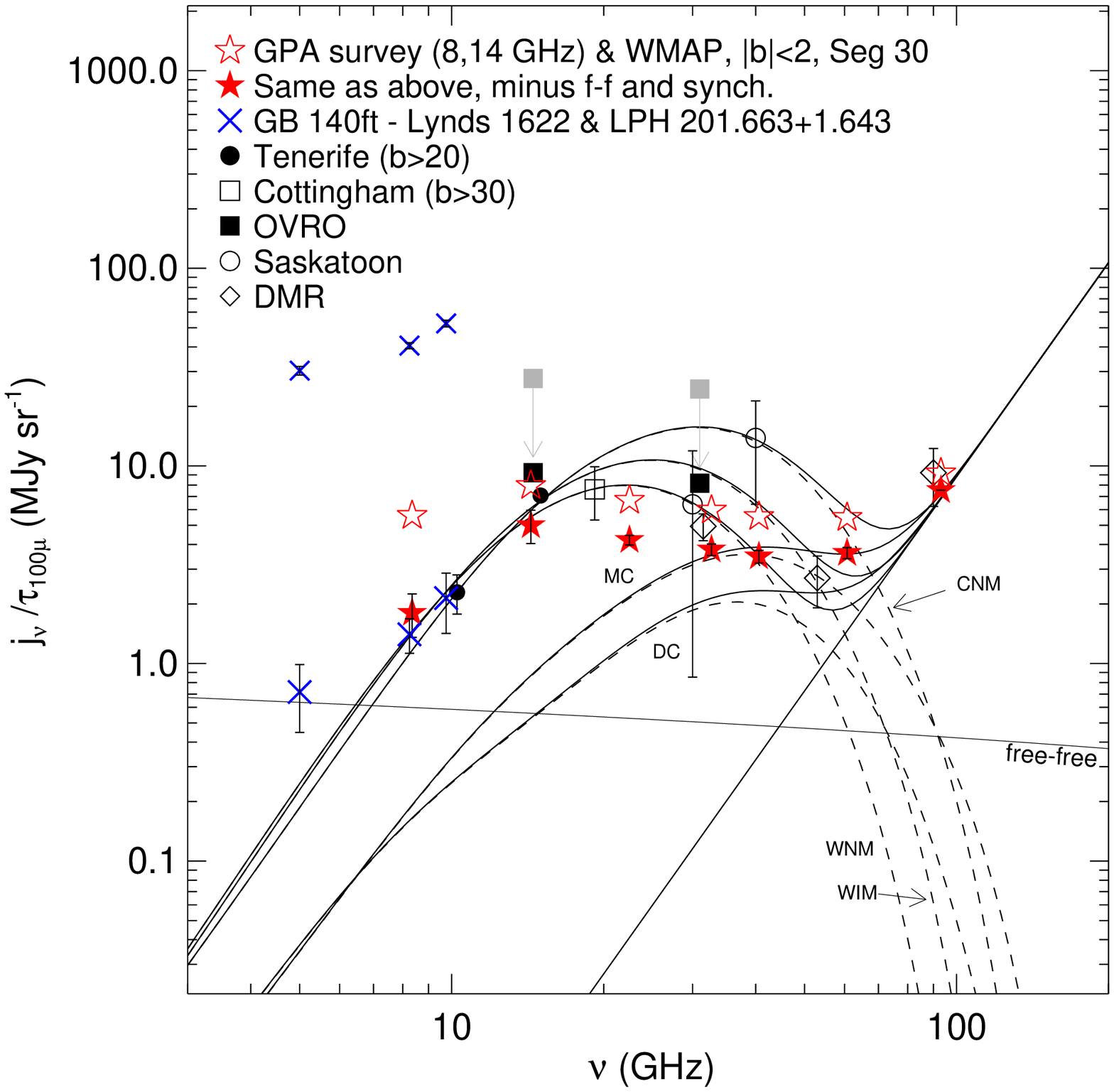}
\caption{Model dust emissivity per $\tau_{100\mu}$ for DC, MC, CNM, WNM, and
WIM conditions (as in Draine \& Lazarian 1998, Figure 9).  Solid thin
lines are total
emissivity; dashed lines are rotational emission. 
Gray line is free-free for
$<n_en_p>/<n_H>=0.01\cm^{-3}$ averaged along the line of sight. 
Also shown are measurements from the \COBE/DMR (\emph{open diamonds})
from Finkbeiner \etal (1999), similar to Kogut
\etal (1996); Saskatoon (\emph{open circles}) (de Oliveira-Costa 
\etal 1997); the Cottingham \& Boughn $19.2\GHz$ survey 
(\emph{open square}) (de
Oliveira-Costa \etal 1998), OVRO data (\emph{solid squares}) 
(Leitch \etal 1997);
Tenerife data (\emph{solid circles})(de Oliveira-Costa \etal 1999);
GB140 data for \LPH\ (\emph{upper crosses}) and LDN1622 (\emph{lower crosses};
Finkbeiner \etal 2002); and Green Bank / \WMAP\ data for Segment 30 from
this study with synchrotron and free free included (\emph{open stars})
and removed (\emph{filled stars}) assuming free-free is $10^{-18}$ 
Jy cm$^2$ sr$^{-1}$ H$^{-1}$ at
8.35 GHz and synchrotron is $2.4\times10^{-18}$ Jy cm$^2$
sr$^{-1}$ H$^{-1}$ at 2.326 GHz with 
$\alpha=-1.05$ as in Figure \ref{fig_spectrum4}. 
The OVRO points have been lowered a factor of 3 relative to Draine \&
Lazarian (1998, Figure 9), because the unusual dust temperature near
the NCP caused an underestimate of the H column density along those lines
of sight.  Given the large range of model curves, most measurements 
are consistent with some superposition of spinning dust, vibrational
dust, and free-free emission. 
\label{fig_dl}
}
\end{figure}

\section{PROCESSING STEPS}
\subsection{Segment maps}
The processing begins with the gpa-*raw.fit files on Glen Langston's
website\footnote{http://www.gb.nrao.edu/$^\sim$glangsto/gpa}.  These files contain the
time-ordered data, merely a set of sky coordinates $(l,b)$ and
measurements in X and Ku band, as well as uncertainties.  If the two
bands had the same beam on the sky, it would be possible to compare
them directly, but because the X-band beam is significantly larger
than the Ku-band beam, it is necessary to make a map and
smooth the Ku-band beam before comparison.

Each set of 180 scans (covering $10\degree\times15\degree$) constitute a
``segment.''  Because the scans are not exactly on great circles,
the data are mapped to an intermediate 
scan\#$-b$ map, the
$(n,b)$ grid, to facilitate Fourier destriping.  The instrumental beam
is very well sampled in the scan direction
(samples every $2.4'$) so linear interpolation in $b$ is sufficient to
map to the $(n,b)$ grid.  The image is significantly padded in the $b$
direction and Fourier transformed, revealing significant power near
the $k_b$ axis outside the band limit, representing stripes in the
scan direction.  This power is subtracted, and the inverse Fourier
transform produces a somewhat cleaner $(n,b)$ map.  

After destriping, the data are interpolated along rows of
constant $b$ from $(n,b)$ to $(l,b)$, a Cartesian grid with $2.4'$
pixels.  Again a simple linear interpolation is used, because the
X-band map is sampled well in this direction, and the Ku-band map,
though not quite well sampled, is noisy enough that linear interpolation does
not compromise the data, especially considering the subsequent
smoothing. 

Segments centered on $l=0,15,30,45$ provide the strongest signal for the
current work.  We shall refer to these, and other, segments by their
central longitude hereafter; all segments center on Galactic latitude
zero.  Segment -10 is observed at low elevation and suffers
from troublesome artifacts due to significant ground pick-up coming
through the side-lobes of the antenna.  This raises the system
temperature resulting in ``curved'' features in the (l,b) maps which
are at a constant elevation above the horizon.  These artifacts are
understood and are not a problem for other parts of the sky. 
Segment 60 has weak dust signal and poor
weather.  Segment 75 is dominated by free-free from a beautiful \HII\
region in Cygnus, but
shows little diffuse ISM, and remaining segments in the outer Galaxy
show no significant ISM emission.  Segment 0 is dominated by
free-free, but is included for completeness. 
Therefore, the results in this
paper are derived exclusively from segments 15, 30, and 45. 

\subsection{Sky subtraction}
The total antenna temperature (including system and sky contributions)
for these observations is 68 K at X-band and 83 K at Ku-band (sum of both
polarizations).  The sky contribution to each sample $i$ is $T_{sky,
i} = T_0 + T_{sky}\sec z_i$, where $z_i$ is the zenith angle of each
sample $i$, and $T_0, T_{sky}$ are constants fit for each segment.
This model is adequate in most cases, but in cases where $T_0$ or
$T_{sky}$ vary during the observation of one segment, a roughly
quadratic error can be introduced into the maps, and because X and
Ku-band are observed at the same time, the resulting artifacts will be
highly correlated.  Concerns about this prompted us to re-reduce the
data with a more ``brute force'' approach of fitting a quadratic
function of $b$ (with
$3\sigma$ outlier rejection) to each scan off the plane
($|b|>2\degree$) and subtracting the fit.  The resulting maps for
segments 0,15,30, and 45 are shown in Figure (1-4)a (X-band) and
(1-4)b (Ku-band).  Note that the result of such fitting is to
suppress any extended emission at $|b|>2$ but such emission is
negligible compared to emission in the plane anyway.   
Bright point sources such as W40 (Segment 30) are preserved by
outlier rejection in the baseline fit.  Results in this study use this
quadratic sky subtraction because it is less susceptible to
systematic errors, but the results do not depend significantly on this
choice. 

\subsection{\HII\ region mask}
\label{sec_mask}
The total emission in the Galactic plane is dominated by a few bright
\HII\ regions and supernova remnants.  It is desirable to mask these
out using independent data, such as the \IRAS\ $60/100\micron$ flux
ratio (Beichman \etal\ 1988).  This ratio is approximately $0.2-0.3$
in the diffuse ISM, and 
exceeds $\sim0.4$ in \HII\ regions.  A cut of 0.4 is taken, and grown by 2
pixels in all directions.  This mask is shown in outline in Figure 
\ref{fig_seg00}d.  Results are not very sensitive to this choice. 

\section{Rising spectrum at 8 \& 14 GHz}

The Green Bank data exhibit a rising intensity from 8 to 14 GHz
that cannot be explained by free-free, synchrotron, or thermal
(vibrational) dust emission.  Data for segments 0, 15, 30, and 45 are
shown in Figures 1-4 respectively.  In each case, the X and Ku-band
images are shown, as well as two superpositions of the X and Ku-band
data: one that cancels free-free [$T_K - (14.35/8.35)^{-2.1}T_X$]
and another that cancels the ISM in the Galactic plane outside of the
\IRAS-based \HII\ mask.  Relative calibration between X and Ku-band is
evidently good to better than $\sim10\%$ or the \HII\ regions would
not be canceled by the first superposition so nicely.  Note that
segment 0 (Figure 1) is dominated by free-free emission
and difficult to interpret, but is included for completeness. 

A simple test reveals that the X and Ku-band data are consistent with
spinning dust models and not with free-free or synchrotron emission.
We plot the free-free subtracted [$T_{Ku} - (14.35/8.35)^{-2.1}T_X$]
superposition of the two bands vs. dust column (expressed as
$100\micron$ optical depth, $\tau_{100\mu} = 266\times E(B-V)$
from Schlegel, Finkbeiner \& Davis 1998; hereafter SFD) in
Figure \ref{fig_scatter} and find a
strong positive correlation for segments 15, 30, and 45.  Of course, a
substantial amount of
spinning dust would be canceled in this superposition, but an indirect
comparison to the 5 Draine \& Lazarian (1998) models is possible by
computing the same superposition for each model.  The Draine \&
Lazarian models are expressed as emissivity per H column density, so
we convert $N$(H)
to $100\micron$ optical depth using $2.13\times10^{24}$ H cm$^{-2}$ = unit
$\tau_{100\mu}$ to obtain a slope for each model\footnote{This 
factor assumes a mean dust temperature of 18.175K obtained by SFD98
and a ratio of $8\times10^{21}$ H cm$^{-2}$ per magnitude $E(B-V)$
derived by SFD98 by comparing 21cm \HI\ emission with dust FIR
emission.  This latter factor differs from the value of
$5.8\times10^{21}$ H cm$^{-2}$ reported by Bohlin \etal 1978}.
One symbol is plotted for each
$12'\times12'$ independent region
in each of the four segments; the vast majority fall among the 5
spinning dust model lines.  A region dominated by free-free would fall
on the horizontal line; any emission with a slope steeper than
$\beta=-2.1$ (such as hard synchrotron)
would fall below the line.  A few such regions appear,
such as SNR 3C392 in segment 30, but in general most pixels outside
the \HII\ mask appear to be consistent with the spinning dust
interpretation of Foreground X, and inconsistent with free-free or
synchrotron emission alone. 

\section{Green Bank and \WMAP}
 
The previous section provided support for the spinning dust
hypothesis, by demonstrating a rising spectrum in the Green Bank data
from 8 to 14 GHz.  By combining those data with the Rhodes survey
(2.326 GHz; see Jonas \etal\ 1998) and \WMAP\ ($23-94$ GHz) we can now
produce a spectrum of dust-correlated emission from $2.3-94$ GHz.  

\subsection{Choice of dust template}
By cross-correlating the data in each band against a dust template
(such as SFD) and measuring the correlation slope, we can obtain the
desired information without being confused by sometimes poorly determined
zero points of the various data sets.  As a template we could use SFD
dust (as in the previous section) or dust times some power of dust
temperature (as in Finkbeiner 2004), but for the present
analysis we prefer the \WMAP\ 94 GHz map\footnote{
The effective central frequency for \WMAP\ W-band is $93.1\GHz$ for free-free
and $94.3\GHz$ for dust.  Because thermal dust is the dominant
Galactic emission in W-band, we refer to it as the ``94 GHz''
band throughout.}.
This map is
dominated by thermal dust emission on the Rayleigh-Jeans tail, where
temperature dependence is weak, so it is a good tracer of total dust
column density.  This way the errors in the SFD dust temperature
correction at low Galactic latitude do not propagate into the
analysis.  All three templates have been
used in the following analysis with nearly identical results, but the
94 GHz map produces tighter scatter and small error bars. 

\subsection{Correlation slopes}
\label{sec_corrslope}
Scatter plots for each of 9 channels (0.408, 2.326, 8.35, 14.35, 23,
33, 41, 61, 94 GHz) vs. 94 GHz are shown for segment 30 in Figure
\ref{fig_scatter9} along with best fit linear regression lines.  For this
analysis, all maps are smoothed to $1\degree$ FWHM and the \HII\
region mask is appropriately enlarged.  Dashed
lines indicate the $1\sigma$ width of the distribution of slopes,
\emph{not} the uncertainty in the mean, which is much smaller. 

These slopes are then converted to $j_\nu \tau_{100\mu}^{-1}$ for
comparison with results obtained in the previous section. 
Draine \& Lazarian models refer to $N($H) because they are
physical models
tied to element abundances relative to H.  In practice, the microwave
emissivities  are 
compared to dust optical depth (without regard for whether
the associated H is molecular, neutral atomic, or ionized) and the
models are converted from $j_\nu/n_H$ (Jy cm$^2$sr$^{-1}$ H$^{-1}$) to
$j_\nu \tau_{100\mu}^{-1}$ (Jy sr$^{-1}$)
with a canonical factor  $2.13\times10^{24}$ H
cm$^{-2}$ = unit $\tau_{100\mu}$.  In the case of the 94 GHz template, a
conversion factor of 34.6 mK\footnote{antenna temperature, not
thermodynamic $\Delta T$}
per unit $\tau_{100\mu}$ is applied, corresponding to 0.13 mK per mag
$E(B-V)$ is applied.  This
conversion factor can be fit empirically from the SFD maps and \WMAP\ 94
GHz data, or derived from a comparison
of the FIRAS-based Finkbeiner \etal\ (1999) 94 GHz prediction and
the SFD $\tau_{100\mu}$ map for median dust temperature. 
Correlation slopes are fit for the 4 segments, with segments 15, 30,
and 45 showing a pronounced rise from 8 to 14 GHz and
requiring some sort of anomalous emission (Figure \ref{fig_spectrum4}).  

As a toy model, a superposition of 0.18 times the Draine \& Lazarian
molecular cloud model plus 0.5 times the Draine \& Lazarian WNM
spectrum is used, where the electric dipole moment is increased by a
factor of two for the WNM curve.  The WNM term is an extrapolation
based on the Draine \& Lazarian WNM model for two values of the
electric dipole moment, and is not a rigorous calculation.  This
parameterization should not be taken seriously as a physical
interpretation of the spectrum; it merely shows that a superposition
of reasonable models provides a good fit to the data.  Further study
will be required to fit these curves in detail with a physical
spinning dust model.

\subsection{Magnetic Dust}
Another emission mechanism that could produce Foreground X is magnetic
dust (Draine \& Lazarian 1999; DL99).  This mechanism has nothing to do with
rapid rotation of magnetic dipoles; rather it results from the thermal
fluctuations in the grain magnetization, yielding an additional
thermal emissivity mechanism. 
The published DL99 models have a slow roll off above the peak
intensity, and this roll off results in high theoretical 61 GHz
correlation slopes relative to the \WMAP\ data.  That is to say, for the
measured 61 GHz to be so low, a sharp roll off in the underlying dust
spectrum with increasing frequency is required.  However, it may be
possible to tune the
dust magnetic susceptibility function to produce a steep roll off
(Draine priv. comm.).  We have not explored this parameter space, but
simply note that while the published magnetic models appear to be
ruled out by the data, a variant of such models may not be ruled out
as a majority contribution.  Further theoretical work is
required before the dominant dust-correlated emission mechanism can be
unambiguously identified.  What is certain, however, is that it is not
hard synchrotron emission alone. 

\subsection{Limits on Hard Synchrotron}
The \WMAP\ team (Bennett \etal 2003b) claims that hard synchrotron
($\beta = -2.5$) emission from dusty star forming regions is the
main contributor to dust-correlated emission at 23 GHz.  They
arrived at this conclusion without the benefit of any diffuse ISM measurements
of the Milky Way between 2.3 and 23 GHz.  Now that a rise in the dust
spectrum has been
observed in the 8 and 14 GHz Green Bank data, it is evident that
another emission
mechanism is required.  By assuming that Foreground X
contributes nothing at 8 GHz and rises sharply to 14 GHz (and ignoring
all other components such as free-free) one can make the
conservative statement that hard synchrotron is less than 2/3 of the 14 GHz
intensity, and less than 1/2 at 23 GHz (based on the data in Table 1).
Inclusion of free-free or 
soft synchrotron pushes these limits down further.  We therefore
conclude that dust-correlated hard synchrotron provides less than 1/2
the intensity in the \WMAP\ bands, and possibly much less.

\subsection{Calibration check: W40}
\label{sec_calcheck}
The arguments in this paper depend most heavily on the spectral rise
observed in the 8 and 14 GHz Green Bank maps, and so the calibration
of these maps deserves close scrutiny.  In order to verify the
calibration of the Green Bank data relative to \WMAP\ (which has a
calibration uncertainty of less than 1\%, Hinshaw \etal\ 2003b) we
single out the \HII\ region W40 at $(l,b) = 28.8\degree, 3.5\degree$,
and repeat the above analysis.  W40 is an association of at least 6
luminous stars adjacent to a molecular cloud (see Smith \etal 1985 for
details) and is certainly an inhomogeneous environment.  However, at
the scale of comparison ($1\degree$) it is unresolved.  \WMAP\ 94 GHz
data are again used as a dust template, only all correlation slopes
are multiplied by 1.8 (i.e. 94 GHz emission per $\tau_{100\mu}$ is
high by that factor here, determined by a fit to the W40 spectrum)
to account for the fact that only part of the 94
GHz emission in this \HII\ region is actually thermal dust.  The W40
spectrum (Figure \ref{fig_W40}) appears to be mostly free-free over a
wide spectral range, with a small but significant enhancement above 33
GHz.  The spinning dust component included is the Draine \& Lazarian
molecular cloud model times a factor of 0.15, for reference.  
This is close to the factor of 0.2 suggested by Draine \& Lazarian
to account for the expected depletion of small grains in
molecular clouds. This
plot demonstrates that (at least in segment 30) the Green Bank
calibration (Langston \etal\ 2000) appears to be compatible with \WMAP\
at the $10\%$ level or better.  This indicates that the disagreement
with the hard synchrotron hypothesis is robust, and does not result
from calibration error.

\subsection{Sidelobes}
\label{sec_sidelobes}
\begin{figure}[tb]
\epsscale{0.9}
\plotone{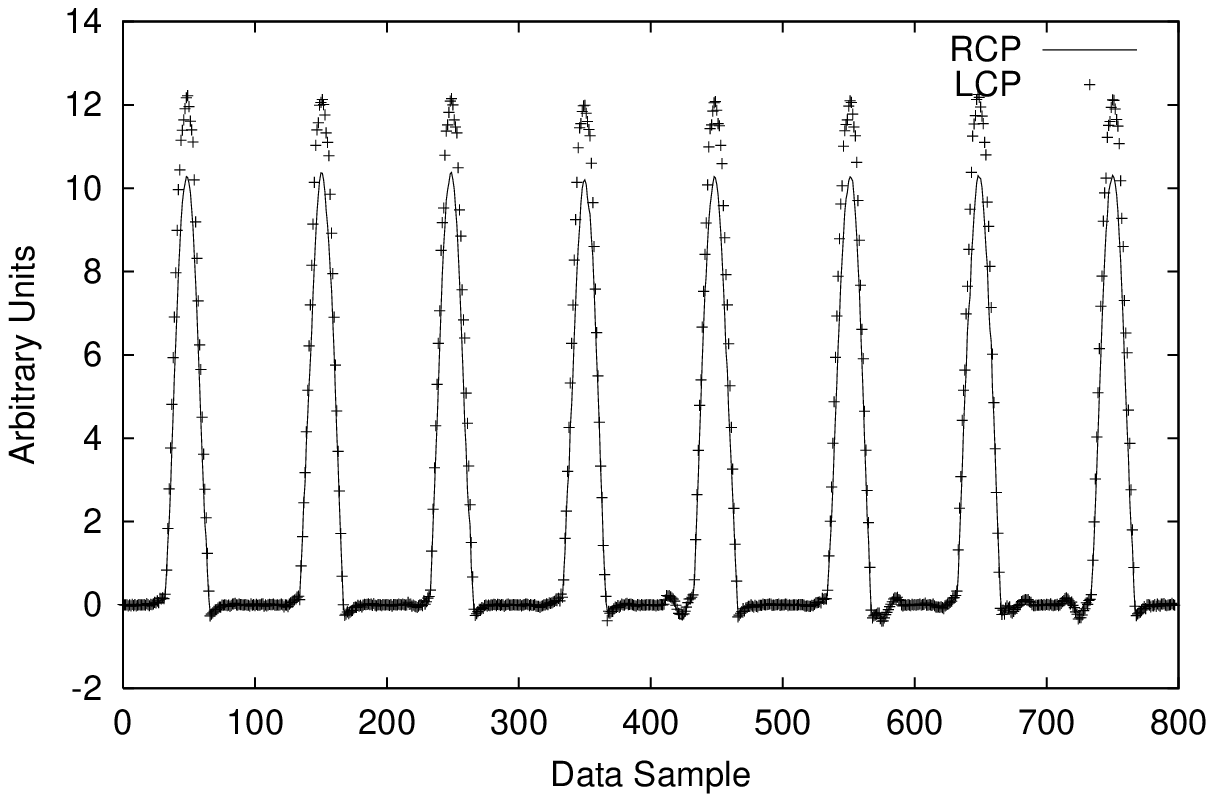}
\figcaption{
\label{fig_sidelobeX}
X-band beam map of Cas A scanning at 8 different approach angles
(every $45\degree$).  Samples are taken every $30''$.
Power in each polarization is shown in
arbitrary units.  Hysteresis may be present at the $2\%$ level, with
other artifacts (diffraction spikes) in some scans at the $3-4\%$
level.  No Airy ring structure is evident at $<1\%$ in
amplitude, meaning $<4\%$ power in the first Airy ring.
}
\end{figure}
\begin{figure}[tb]
\epsscale{0.9}
\plotone{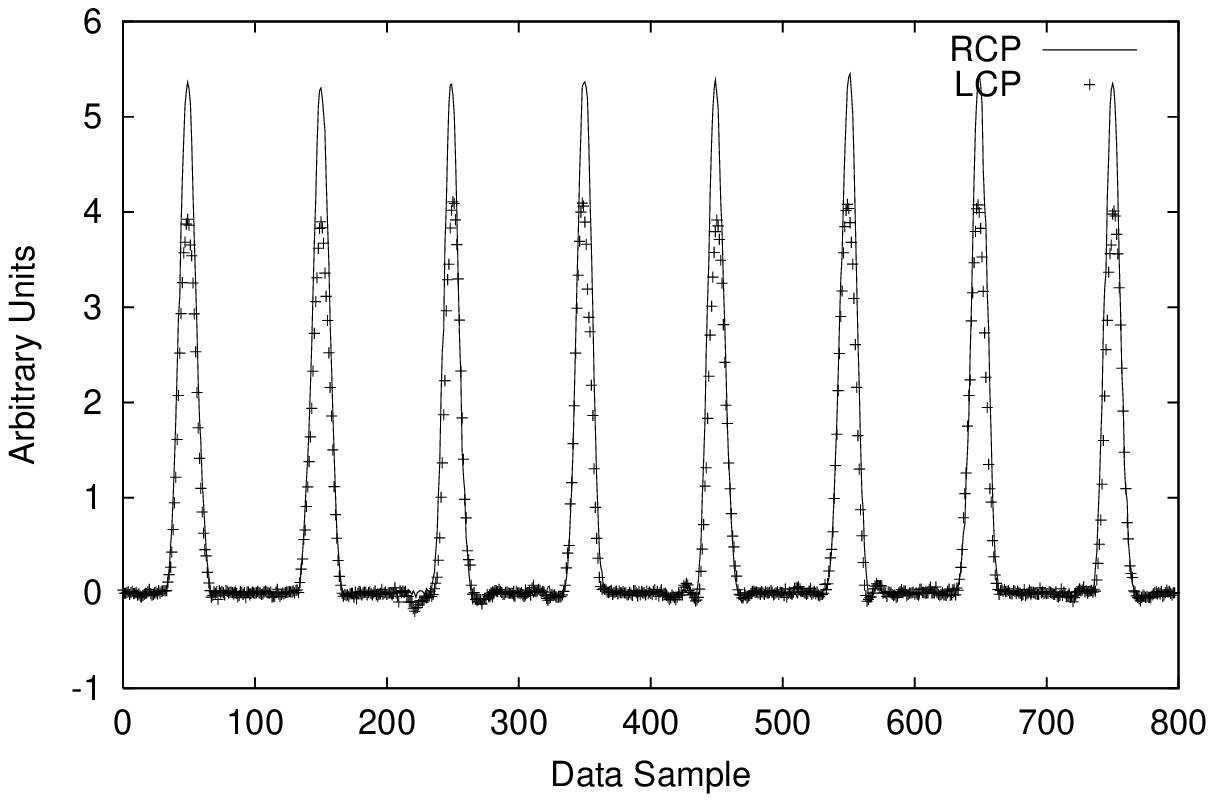}
\figcaption{
\label{fig_sidelobeKu}
Ku-band beam map of Cas A scanning at 8 different approach angles
(every $45\degree$).  Samples are taken every $30''$.
Power in each polarization is shown in
arbitrary units.  The hysteresis seen in X-band is absent, but 
diffraction spikes ($\pm3\%$) appear next to peaks 5 and 6 because
that scan was slightly off center.  A data glitch appears before peak
3 in one polarization.  Sidelobes are estimated to be $<1\%$ in
amplitude, meaning $<4\%$ power in the first Airy ring.
}
\end{figure}

Substantial contribution from ground pickup in beam sidelobes is
always a potential concern with continuum measurements.  At 14 GHz,
the brightest regions on the sky are of order 1 K and the ground is
300K, so ground emission is far more important than reflection of the
sky.  Such ground emission does appear in the
segment centered at $l=-10$, creating an obvious constant elevation
pattern, so this segment has been discarded. 

Far sidelobes on the sky are not a concern, because the sky baseline
for each scan is established by fitting each scan at the ends, $|b| >
2\degree$, with a quadratic fit.  The far (tens of degrees) sidelobes
can only produce a very smooth background, and are subtracted by this
fit.  The only potential contamination comes from the near
sidelobes (few degrees off axis) picking up bright objects on the sky.

To verify that near sidelobes are negligible, we scanned the
unresolved source Cas A from 8 approach angles, i.e. every by
$45\degree$ (Figures \ref{fig_sidelobeX}, \ref{fig_sidelobeKu}).
Artifacts away from the main beam, such as diffraction spikes, are at
the few percent level.  There
is no sign of the second Airy ring, which for a 2m secondary mirror
blocking the 13.7m primary would theoretically have an amplitude of
2.4\% of the peak and contain 10\% of the beam power.  Where the scans
are slightly off center, diffraction spikes of 3\% amplitude can be
seen in Fig. \ref{fig_sidelobeKu}) near peaks 5 and 6, so the unseen
sidelobes are probably less than 1\%.  This happens because the
illumination of the primary is not uniform, but rather rolls off to
near zero at the edge of the dish.  This apodization suppresses the
airy rings, making the amplitude less than the theoretical 2.4\%.  A
1\% amplitude (which seems conservative) implies 4\% or less of the
total power is scattered into the first Airy ring.  Unfortunately,
this telescope is no longer in service, so more complete beam maps are
impossible to obtain. 

It is difficult to imagine sidelobe contamination affecting
our correlation slopes by more than a few percent, compared with the
factor of 2 discrepancy between the X- and Ku-band data and the
Bennett \etal\ synchrotron model.

\section{SUMMARY}

A previous analysis (Finkbeiner 2004) showed that \WMAP\ data off the
Galactic plane are consistent with spinning dust emission similar to
that proposed by Draine \& Lazarian (1998), but agreed with Bennett
\etal (2003b) that dust-correlated hard synchrotron ($\beta\sim-2.5$) 
could explain the emission as well.  \WMAP\ alone
cannot differentiate between these two possibilities because the
spectral rise predicted for most spinning dust models occurs at lower
frequencies ($\nu < 23 $ GHz).  Because there are no other data of
comparable quality at high latitude in the required frequency range,
this study makes use of the Green Bank Galactic Plane Survey (8.35,
14.35 GHz) at low latitude ($ -4 < b < 4\degree$), finding a
significant spectral rise and ruling out hard synchrotron as a majority
contributor at 23 GHz.  In light of this discovery, it is likely that
Foreground X dominates hard synchrotron at higher latitudes as well.

Previous attempts to understand microwave foregrounds have missed
important features in the data (e.g. by ignoring Foreground X
altogether, ignoring effects of dust temperature variation, or failing to
identify the free-free Galactic haze seen by Finkbeiner 2004) and we
may continue to oversimplify by assuming the same physical mechanism
is responsible for all occurrences of Foreground X.  In this early
phase of the investigation, we have attempted to keep the 
analysis described herein (fitting correlation slopes as a
function of frequency) as straightforward as possible,
aiming to demonstrate the existence of Foreground X and motivate
further measurements.  More sophisticated techniques
may be used in the future to refine our understanding of Foreground(s)
X, and to derive parameters of physical models for the
emission (as in  Draine \& Lazarian).  For example, polarization
measurements of clouds such as LDN1622 at $3-10\GHz$ may constrain
spinning dust models (see Lazarian \& Draine 2000). 
The present analysis simply argues
strongly for the existence of spinning dust or another component with
similar spectral behavior. 


We are indebted to Bruce Draine and David Schlegel for encouragement and
advice.  
Carl Heiles and Angelica de Oliveira-Costa provided helpful
conversations. 
This research made use of the NASA Astrophysics Data
System (ADS) and the The IDL Astronomy User's Library at
Goddard Space Flight
Center\footnote{http://idlastro.gsfc.nasa.gov/}.  DPF is a Hubble
Fellow supported by HST-HF-00129.01-A and by 
NASA LTSA grant NAG5-12972.  The Green Bank Earth Station
telescope is located at Green Bank Observatory, operated by the
National Radio Astronomy Observatory (NRAO), which is a facility of
the National Science Foundation operated under cooperative agreement
by Associated Universities, Inc.
\newpage

\bibliographystyle{unsrt}
\bibliography{gsrp}


\clearpage
\begin{deluxetable}{r|rrrrr}
\footnotesize
\tablewidth{0pt}
\tablecaption{Correlation slope results
   \label{table_results}
}
\tablehead{
\colhead{$\nu$} &
\colhead{Seg  0} &
\colhead{Seg 15} &
\colhead{Seg 30} &
\colhead{Seg 45} &
\colhead{W40}
\\
\colhead{GHz} &
\colhead{$j_\nu / \tau_{100\mu}$} &
\colhead{$j_\nu / \tau_{100\mu}$} &
\colhead{$j_\nu / \tau_{100\mu}$} &
\colhead{$j_\nu / \tau_{100\mu}$} &
\colhead{$j_\nu / \tau_{100\mu}$}}
\startdata
 0.408 &$18.30 \pm 5.60 $&$18.75 \pm 5.71 $&$11.80 \pm 3.52 $&$13.37 \pm 4.23 $&$11.33 \pm 5.00 $ \\
 2.326 &$10.33 \pm 1.36 $&$ 9.81 \pm 1.31 $&$ 7.90 \pm 0.96 $&$ 7.36 \pm 1.14 $&$17.68 \pm 1.16 $ \\
 8.35~ &$ 5.37 \pm 0.51 $&$ 5.13 \pm 0.57 $&$ 5.60 \pm 0.44 $&$ 4.67 \pm 0.61 $&$12.56 \pm 1.18 $ \\
14.35~ &$ 5.38 \pm 0.99 $&$ 7.75 \pm 1.34 $&$ 7.95 \pm 0.95 $&$ 7.62 \pm 1.05 $&$12.61 \pm 1.14 $ \\
22.5~~ &$ 6.05 \pm 0.43 $&$ 6.65 \pm 0.25 $&$ 6.69 \pm 0.23 $&$ 6.15 \pm 0.24 $&$12.59 \pm 0.41 $ \\
32.7~~ &$ 5.24 \pm 0.39 $&$ 5.86 \pm 0.21 $&$ 5.97 \pm 0.25 $&$ 5.19 \pm 0.24 $&$12.87 \pm 0.39 $ \\
40.6~~ &$ 4.86 \pm 0.36 $&$ 5.42 \pm 0.20 $&$ 5.53 \pm 0.26 $&$ 4.70 \pm 0.23 $&$12.72 \pm 0.43 $ \\
60.7~~ &$ 4.81 \pm 0.33 $&$ 5.35 \pm 0.18 $&$ 5.47 \pm 0.23 $&$ 4.69 \pm 0.21 $&$12.37 \pm 0.36 $ \\
93.1~~ &$ 9.20 \pm 0.00 $&$ 9.20 \pm 0.00 $&$ 9.20 \pm 0.00 $&$ 9.20 \pm 0.00 $&$16.57 \pm 0.00 $ \\
\enddata
\tablecomments{Correlation slopes (emission per dust $100\micron$
optical depth, $\tau_{100\mu}$) for Haslam (0.408 GHz), Rhodes
(2.326 GHz), Green Bank (8.35, 14.35 GHz), and \WMAP\ ($22.5-93.1$ GHz)
in units of MJy sr$^{-1}$.  To convert to emissivity per H atom (Jy
cm$^2$sr$^{-1}$ per H), divide by $2.13\times10^{18}$.  
Error values represent the $1\sigma$ width of the distribution of slopes as seen in
Figure \ref{fig_scatter9}, not the (much smaller) uncertainty of the mean.
The \WMAP\ bands
are rather broad; the listed \WMAP\ frequencies are effective central
frequencies for emission with a free-free spectrum, given by Page \etal
(2003).  The highest frequency \WMAP\ band, dominated by thermal dust
emission, is used as the ``dust template'' in this paper, and
has no measured error.  
The W40 spectrum is substantially contaminated by free-free emission
at 94 GHz, resulting in a higher normalization
(see \S\ref{sec_calcheck} for details). 
Conversion from \WMAP\ thermal dust to
$\tau_{100\mu}$ is described in \S\ref{sec_corrslope}.
}
\end{deluxetable}
\clearpage

\end{document}